\documentclass[12pt]{article}
\usepackage{latexsym}
\usepackage{amsfonts}
\usepackage{epsfig}

\catcode`\@=11

\def\section{\@startsection{section}{1}{\z@}{3.5ex plus 1ex minus
 .2ex}{2.3ex plus .2ex}{\bf}}

\def\thesubsection{\arabic{section}.\arabic{subsection}}
\renewcommand{\subsection}[1]{\addtocounter{subsection}{1}
\vspace{2.5mm}\par\noindent {\it \thesubsection . #1}\par
 \vspace{0.5mm} }

\catcode`\@=12
\newfont{\mbm}{msbm10 scaled\magstep1}

\mathchardef\varGamma="0100 \mathchardef\varDelta="0101
\mathchardef\varTheta="0102 \mathchardef\varLambda="0103
\mathchardef\varXi="0104 \mathchardef\varPi="0105
\mathchardef\varSigma="0106 \mathchardef\varUpsilon="0107
\mathchardef\varPhi="0108 \mathchardef\varPsi="0109
\mathchardef\varOmega="010A

\DeclareFontFamily{U}{rsf}{}
\DeclareFontShape{U}{rsf}{m}{n}{
  <5> <6> rsfs5 <7> <8> <9> rsfs7 <10-> rsfs10}{}
\DeclareMathAlphabet\Scr{U}{rsf}{m}{n}

\begin{document}
\begin{titlepage}
\rightline{SU-ITP-04/01}
\rightline{{hep-th/0402047}}
\vskip 2cm
\centerline{{\large\bf Volume Stabilization and  the Origin of the Inflaton
}}
\vskip 0.3cm
\centerline{{\large\bf  Shift Symmetry in String Theory}}
\vskip 1cm
\centerline{\bf Jonathan P. Hsu and Renata Kallosh}
\vskip 0.3cm

\centerline{\it  pihsu@stanford.edu, \, kallosh@stanford.edu
}

\vskip 0.3cm

\centerline{\it Department of Physics, Stanford University, Stanford, CA 94305}
\vskip  1.0cm
\begin{abstract}

One of the main problems of inflation in string theory is finding
models with a flat potential while simultaneously stabilizing the
volume of the compactified space. This can be achieved in theories
where the potential has (an approximate) shift symmetry in the
inflaton direction. We will identify a class of models where the
shift symmetry uniquely follows from the underlying mathematical
 structure of the
theory. It is related to the symmetry properties of the
corresponding coset space and the period matrix of special
geometry, which shows how the gauge coupling depends on the volume
and the position of the branes. In particular, for type IIB string
theory on $K3\times T^2/\mathbb{Z}_2$  with  D3 or D7 moduli
belonging to vector multiplets, the shift symmetry is  a part of
$SO(2,2+n)$ symmetry of the coset space  $\left ({SU(1,1))\over
U(1)}\right)\times {SO(2,2+n)\over (SO(2)\times SO(2+n)}$.  The
absence of a prepotential, specific for the stringy version of
supergravity, plays a prominent role in this construction, which
may provide a viable mechanism for the accelerated expansion and
inflation in the early universe.

\end{abstract}
\end{titlepage}

\section{Introduction}

One of the  problems of string cosmology is to provide a
 mechanism for the accelerated expansion and inflation in
the early universe, which would be consistent with stabilization
of the volume of the compactified space. A possible solution of
this problem for the accelerated expansion in a metastable dS
space was recently proposed in \cite{KKLT}. However, in order to
generalize this solution for the usual slow-roll inflation one
would need to find a potential containing a flat direction for the
inflaton field. One would also need to make sure that the motion
of the field in this potential does not destabilize the volume
\cite{KKLMMT}.

It has been suggested in \cite{KalloshHsuProk} that flat
directions for the inflaton field in the D3/ D7 brane inflation
model \cite{DHHK}, consistent with the volume stabilization, can
appear as a consequence of shift symmetry with respect to the
inflaton field. The existence of this symmetry follows from the
assumption of the existence of the BPS state of branes with
unbroken supersymmetry. More recently, it was argued in
\cite{Firouzjahi:2003zy} that under the same assumption, namely,
the existence of unbroken supersymmetry of the BPS state of
branes, the shift symmetry might also appear in the
$D3-\overline{D3}$ inflation model of \cite{KKLMMT}.

The goal of this paper is to show that in a certain class of
string theory models the shift symmetry, as well as the existence
of the supersymmetric BPS ground state, is not an assumption but
an unavoidable consequence of the underlying mathematical
structure of the theory. We will give an example of a class of
models where this is indeed the case and show that the shift
symmetry and the flatness of the inflaton potential in these
models are related to the symmetry properties of the corresponding
coset space, and the period matrix, which shows how the gauge
coupling depends on the volume and the position of the branes.

We will illustrate our general approach investigating the  D3/ D7
system \cite{DHHK} of type IIB string theory. We will study it in
the context of the special geometry construction
\cite{Angelantonj:2003zx}, \cite{KTW}. We will be interested here
in the setting \cite{KalloshHsuProk}  where the supersymmetry
breaking effects, like non-self-dual fluxes on D7 brane, are not
yet included. We would like to learn how the stabilization of
volume affects the  state of the branes, whether their position
can still be a modulus. We will see that for D3/ D7 system  {\it
one inevitably finds an effective theory with stabilization of the
volume and a shift symmetry for the inflaton}. This means that
these branes (D3 or D7) move freely in the theory with volume
stabilization. A small deviation from the shift symmetry will lead
to a slow-roll inflation.

Our starting point is the construction presented  by Angelantonj,
D'Auria, Ferrara, Trigiante (ADFT) in \cite{Angelantonj:2003zx}
\footnote{In \cite{KTW} a model with only D3 moduli and with
cosmological applications was proposed. A corrected consistent
version of it is closely related to the one in
\cite{Angelantonj:2003zx}.}. They have extended the
four--dimensional gauged supergravity analysis of type IIB vacua
on $K3\times T^2/\mathbb{Z}_2$ to the case where the D3 and D7
moduli, belonging to ${\cal N}=2$ vector multiplets, are turned
on. We will be interested in two cases when either D3 or D7 moduli
are present. In each case the overall special geometry corresponds
to a symmetric space. We will specify the fundamental shift
symmetry for  the D7  moduli  as a block diagonal symplectic
matrix and for the D3 moduli as a lower triangular block form of a
symplectic matrix. They will present a part of $SO(2,2+n)$
symmetry of the coset space  $\left ({SU(1,1))\over
U(1)}\right)\times {SO(2,2+n)\over (SO(2)\times SO(2+n)}$. We will
identify  the gauge couplings in these theories, which will lead
us to particular choices of the non-perturbative superpotentials
to be used for the stabilization of the volume of the internal
space. As the result, in both cases, the potential after volume
stabilization respects the inflaton shift symmetry.

\subsection{ On special K\"ahler geometry}

%%%%%%%%%%%%%%%%%%%%%%%%%%%%%%%%%%
Let us briefly recall the main formulae of special K\"ahler
geometry  \cite{deWit:1984pk,Andrianopoli:1996cm}. The geometry of the manifold
is encoded in the holomorphic section
$\varOmega=(X^\varLambda,\,F_\varSigma)$ which, in the {\it
special coordinate} symplectic frame, is expressed in terms of a
prepotential ${\cal
F}(s,t,u,x^k,y^r)=F(X^\varLambda)/(X^0)^2={\cal
F}(X^\varLambda/X^0)$, as follows:
\begin{equation}
\varOmega = ( X^\varLambda,\,F_\varLambda=\partial F/\partial
X^\varLambda)\,.\label{specialcoordinate}
\end{equation}
In our case ${\cal F}$ is given by Eq. (\ref{prepot}). The
K\"ahler potential $K$ is given by the symplectic invariant
expression:
\begin{equation}
K = -\log \left[{\rm i}(\overline{X}^\varLambda
F_\varLambda-\overline{F}_\varLambda X^\varLambda)\right] \,.
\end{equation}
Symplectic transformations are realized both on the section
$\varOmega$ as well as on vector fields $({\cal
F}_{\mu\nu}^{-\Lambda}, G^{\mu\nu}_{-\Lambda})$:
\begin{equation}
\left(\matrix{X^\Lambda  \cr F_\Lambda }\right)'= \left(\matrix{A
& -B\cr C & D}\right)\left(\matrix{X^\Lambda  \cr F_\Lambda
}\right) \,. \label{duality}\end{equation} The matrix ${\cal
S}=\left(\matrix{A & -B\cr C & D}\right)$ is an $Sp\left( 2(4+k),
\mathbb{R}\right)$ matrix, with
\begin{equation}A^T C-C^T A=0\ \ ,\ \ B^T D- D^T B=0\ \ , \ \  A^T D-C^T B=I
 \end{equation}
In terms of $K$ the metric has the form
$g_{i\bar{\jmath}}=\partial_i\partial_{\bar{\jmath}}K$. The period
matrix ${\cal N}$ defines the vector kinetic lagrangian as
follows:
$$
{\rm Im } \, {\cal F}^{-\Lambda}_{\mu\nu} {\overline{ \cal
N}}_{\Lambda \Sigma}   {\cal F}^{-\Sigma \mu\nu }= -2 {\rm Im }\,
{\overline{ \cal N}}_{\Lambda \Sigma} \, {\cal
F}^{\Lambda}_{\mu\nu}\, {\cal F}^{\Sigma \mu\nu} + {\rm Re}\,
{\overline{ \cal N}}_{\Lambda \Sigma}\, {\cal
F}^{\Lambda}_{\mu\nu} \, \tilde{{\cal F}}^{\Sigma \mu\nu}
$$
The period matrix  can be introduced via the relations
\begin{equation}
\overline {F_{\Lambda}} = {\overline{ \cal N}}_{\Lambda \Sigma}
\overline {X^{\Sigma}}\, , \qquad h_{\Lambda | i} = {\overline{
\cal N}}_{\Lambda \Sigma} f^\Sigma _i
\end{equation}
\begin{eqnarray}
\overline{{\cal N}}_{\varLambda\varSigma}&=& \hat{h}_{\varLambda|I}\circ
(\hat{f}^{-1})^I{}_\varSigma\,,\,\,\mbox{where}\,\,\,\,
\hat{f}_{I}^\varLambda = \left(\matrix{{\cal D}_i X^\varLambda \cr
\overline{X}^\varLambda }\right)\,;\,\,\,\,\hat{h}_{\varLambda|
I}=\left(\matrix{{\cal D}_i F_\varLambda \cr \overline{F}_\varLambda
}\right) \,.
\end{eqnarray}
The equations of motion  of ${\cal N}=2$ supergravity are
invariant under symplectic transformations (also the action,
except for the kinetic term for the vectors) if the period matrix
transforms as follows
\begin{equation}
({ \cal N}) ' = (C+D { \cal N})(A+B { \cal N})^{-1}
\label{periodtrans}
\end{equation}
These transformations are known as dualities. Only part of these
symmetries can be realized in perturbation theory as the symmetry
of the action: this requires a lower triangular block form of the
symplectic matrix with $B=0$, $D=(A^T)^{-1}$, $A^T C-C^T A=0$:
\begin{equation}
  {\cal S}_{pert}= \left(\matrix{A & 0\cr C & (A^T)^{-1}}\right)
\label{triangular}
\end{equation}
Under these transformations electric fields ${\cal
F}_{\mu\nu}^\Lambda$  are not mixed with magnetic ones ${\cal
G}_{\Lambda \mu\nu}$ and $X^\Lambda$ are not mixed with
$F_{\Lambda}$. The lagrangian under such change is invariant up to
a surface term
\begin{equation}
  {\cal L}' = {\cal L} + {\rm Im}[ (C^T A)_{\Lambda \Sigma}\,
  {\cal F}^{-\Lambda}_{\mu\nu}\, {\cal F}^{-\Sigma \mu\nu }]=
  {\rm Re}[ (C^T A)_{\Lambda \Sigma}]\, {\cal F}^{\Lambda}_{\mu\nu}\,
  \tilde{{\cal F}}^{\Sigma \mu\nu }
\label{L}
\end{equation}
It has been discovered in \cite{CDFV} that heterotic string theory
requires a version of special geometry which is based on the
symplectic section for which the prepotential does not exist. In
standard supergravities where the prepotential exist, $X^\Lambda(
t^i)$ depend on all independent special coordinates $ t^i$. Here
$\Lambda = 0, 1, ..., n$ and $i=1,2, ..., n$. To define the
no-prepotential case we have to split all special coordinates into
a group $t^i= (t^1=s, t^a)$ where $a=2, ..., n$.
A prepotential only exists
when the upper part of the symplectic section can be invertibly
mapped to the coordinates of the special K\"ahler manifold. When
this is not the case, i.e. when 
\begin{equation}
\partial X^\varLambda /\partial s =0
\label{noprep}
\end{equation}
for one of the coordinates
$s$, then no prepotential exists.
 However, a completely consistent version of stringy
${\cal N}=2$ supergravity based on the symplectic section is
available. In the heterotic theory $s$ is a dilaton-axion
superfield, in type IIB it is a volume-4-form superfield.  The
no-prepotential theory has some particular features which will be
used heavily in what follows.

\section{ Overview of ADFT construction}

%%%%%%%%%%%%%%%%%%%%%%%%%%%%%%%%%%%%
In the absence of open--string moduli the four--dimensional ${\cal
N}=2$ effective supergravity is defined by a special geometry
which is described by the coset space $ \label{stu}
    \left(\frac{{\rm SU} (1,1)}{{\rm U}(1)}\right)_s\times
\left(\frac{{\rm SU}(1,1)}{{\rm U}(1)}\right)_t\times
    \left(\frac{{\rm SU}(1,1)}{{\rm U}(1)}\right)_u\,,
$
where $s,\,t,\,u$ denote the scalars of the vector multiplets
containing the $K3$--volume and the R--R $K3$--volume--form, the
$T^2$--complex structure, and the IIB axion--dilaton system,
respectively:
\begin{eqnarray}
s= C_{(4)} -{\rm i}\, {\rm Vol} (K_3),\,
\qquad
t= \frac{g_{12}}{g_{22}} +{\rm i}\,\frac{\sqrt{{\rm det}
g}}{g_{22}}\,,
\qquad
u= C_{(0)} +{\rm i}\, e^{\phi}\,,
\end{eqnarray}
where the matrix $g$ denotes the metric on $T^2$. The total volume
of $T^2$, $\sqrt{{\rm det} g}$ belongs to the hypermultiplet and
will not be considered here since we will first focus on  special
geometry.

It was explained in ADFT that the system of interest can be described
starting from the following unique trilinear prepotential of special
geometry:
\begin{equation}\label{prepot}
    {\cal F}(s,t,u,x^k,y^r)\,=\, stu-\frac{1}{2}\,s \,x^k
    x^k-\frac{1}{2}\,u\,
    y^r y^r\,,
\end{equation}
where $x^k$ and $y^r$ are the positions of the D7 and D3--branes
along $T^2$ respectively, $k=1,\dots, n_7$, $r=1,\dots ,n_3$, and
summation over repeated indices is understood. The prepotential in
Eq. (\ref{prepot}) corresponds to the homogeneous not symmetric
spaces called $L(0,n_7,n_3)$ in \cite{van}.  If we set either all the $x^k$ or all the
$y^r$ to zero, the special geometry in each case describes a
symmetric space:
\begin{eqnarray}\label{symanifold7}
\left(\frac{{\rm SU}(1,1)}{{\rm U}(1)}\right)_s\times
\frac{{\rm SO}(2,2+n_7)}{{\rm SO}(2)\times
{\rm SO}(2+n_7)}\,,\,\,\,\,\,\,\mbox{for
$y^r=0$}\,,\\\label{symanifold3}
\left(\frac{{\rm SU}(1,1)}{{\rm U}(1)}\right)_u\times
\frac{{\rm SO}(2,2+n_3)}{{\rm SO}(2)\times
{\rm SO}(2+n_3)}\,,\,\,\,\,\,\,\mbox{for $x^k=0$}\,.
\end{eqnarray}
 The components $X^\varLambda,\,F_\varSigma$ of
the symplectic section which correctly describe our problem, are
chosen by performing a constant symplectic change of basis from
the one in (\ref{specialcoordinate}) given in terms of the
prepotential in Eq. (\ref{prepot}). The rotated symplectic section
is given by
\begin{tabbing}
\hskip 11 cm \= \kill \\
\hskip 2 cm $X^0 = \frac{1}{{\sqrt{2}}}\,(1 - t\,u +
\frac{(x^k)^2}{2})\,,$\,\> $F_0 = \frac{s\,\left( 2 - 2\,t\,u +
(x^k)^2 \right) +
u\,(y^r)^2}{2\,{\sqrt{2}}}$\,,\\
\hskip 2 cm $X^1 = -\frac{t + u}{{\sqrt{2}}}$\,,\>$F_1 =
  \frac{-2\,s\,\left( t + u \right)  +
  (y^r)^2}{2\,{\sqrt{2}}}$\,,\\
\hskip 2 cm $X^2 =  -\frac{1}{{\sqrt{2}}}\,({1 + t\,u -
\frac{(x^k)^2}{2}})$\,,\>$F_2=
  \frac{s\,\left( 2 + 2\,t\,u - (x^k)^2 \right)  -
  u\, (y^r)^2}{2\,{\sqrt{2}}}$\,,\\
\hskip 2 cm $X^3 = \frac{t - u}{{\sqrt{2}}}$\,,\>$F_3 = \frac{2\,s\,\left( -t + u \right)  + (y^r)^2}{{2\,\sqrt{2}}}$\,,\\
\hskip 2 cm $X^k = x^k$\,,\>$F_k = -
  s\,x^k$\, ,\\
\hskip 2 cm $X^r = y^r$\,,\>$F_r = -u\,y^r$\,.\\
\end{tabbing}
\vskip -0.7 cm
The  K\"ahler potential $K$ has the following form:
\begin{equation}
K = -\log\left[-8\,({\rm Im}(s)\,{\rm Im}(t){\rm
Im}(u)-\frac{1}{2}\,{\rm Im}(s)\,({\rm
Im}(x)^i\,)^2-\frac{1}{2}\,{\rm Im}(u)\,({\rm
Im}(y)^r\,)^2)\right] \,,
\end{equation}
 with ${\rm Im}(s)<0$ and ${\rm Im}(t),\,{\rm
Im}(u)>0$ at $x^k=y^r=0$.
Note that, since $\partial X^\varLambda /\partial s =0$ the new sections do
not admit a prepotential, and the no--go theorem on partial supersymmetry
breaking \cite{Cecotti} does not apply in this case.

\section{ Shift symmetries in positions of  D7 branes}

%%%%%%%%%%%%%%%%%%%%%%%%%%%%%%%%%%
The  symplectic section and the K\"ahler potential in absence of D3 moduli are given by
\begin{tabbing}
\hskip 11 cm \= \kill \\
\hskip 2 cm $X^0 = \frac{1}{{\sqrt{2}}}\,(1 - t\,u +
\frac{(x^k)^2}{2})\,,$\,\> $F_0 = \frac{s\,\left( 2 - 2\,t\,u +
(x^k)^2 \right)
}{2\,{\sqrt{2}}}$\,,\\
\hskip 2 cm $X^1 = -\frac{t + u}{{\sqrt{2}}}$ \,,\>$F_1 =
  \frac{-2\,s\,\left( t + u \right)
  }{2\,{\sqrt{2}}}$\,,\\
\hskip 2 cm $X^2 =  -\frac{1}{{\sqrt{2}}}\,({1 + t\,u -
\frac{(x^k)^2}{2}})$\,,\>$F_2=
  \frac{s\,\left( 2 + 2\,t\,u - (x^k)^2 \right)
  }{2\,{\sqrt{2}}}$\,,\\
\hskip 2 cm $X^3 = \frac{t - u}{{\sqrt{2}}}$\,,\>$F_3 = \frac{2\,s\,\left( -t + u \right)  }{{2\,\sqrt{2}}}$\,,\\
\hskip 2 cm $X^k = x^k$\,,\>$F_k = -
  s\,x^k$\, .\\
\end{tabbing}
\vskip -0.7 cm
\begin{equation}
K = -\log[-8\,{\rm Im}(s)] - \log [{\rm Im}(t){\rm
Im}(u)+\frac{1}{16}\,({\rm Im}(x)^i\,)^2] \,.
\end{equation}
The $SO(2, 2+n_7)$ symmetry is realized manifestly. It is useful
to switch to  light-cone variables, $ X^{\mp}= {X^0\mp X^2\over
\sqrt{2}}$ and $ Y^{\mp}= {X^1\mp X^3\over \sqrt{2}}, $ where
\begin{equation}
 X^- X^+ + X^+ X^- + Y^- Y^+ + Y^+ Y^-- (X^k)^2=0
\label{surface}
\end{equation}
In this basis
\begin{equation}
X^\Lambda =\{  X^- = 1, \,  X^+ = -tu +x^2/2, \, Y^- = -t, \,
Y^+ = -u,\,  X^k = x^k \} \end{equation}
\begin{equation}F_\Lambda =\{  F_-^X = s(-tu +x^2/2), \,  F_+^X = s, \, F^Y_- = -su,
\, F_+^Y = -st,\,
  F_k = -sx^k \} \end{equation}
Thus the model is defined by
\begin{equation}
X^\Lambda \eta_{\Lambda \Sigma} X^\Sigma=0 \qquad
F_\Lambda = s \eta_{\Lambda \Sigma} X^{\Lambda}
\label{SOF}
\end{equation}
Where indices are raised and lowered using $\eta$
\begin{eqnarray}
&&\eta\,=\,\,\left(\matrix{0 &1&0&0&0\cr 1
&0&0&0&0\cr 0 &0&0&1&0\cr 0 &0&1&0&0\cr 0 &0 &0&0&-1}\right)\,.
\end{eqnarray}
The shift of the D7 moduli with the real parameters $\alpha^k$ is
given by,
\begin{equation}
 (x^k )' = x^k +\alpha^k \qquad s'=s \qquad t'=t \qquad u'=u
\label{D7shift}
\end{equation}
can be realized as a perturbative symplectic transformation, a
block-diagonal matrix
$${\cal S}^{D7} = \left(\matrix{A & 0\cr 0 & (A^T)^{-1}}\right)$$
with $B=C=0$ and $D= (A^T)^{-1}$. For the simple case of one D7
brane, $k=1$,  $A^{\Lambda}{}_{ \Sigma}$  is equal to
\begin{eqnarray}
&&A\,=\,\,\left(\matrix{1 &0&0&0&0\cr {\alpha^2\over 2}
&1&0&0&\alpha\cr 0 &0&1&0&0\cr 0 &0&0&1&0\cr \alpha &0 &0&0&1}\right)\,,
\end{eqnarray}
so that the shift is defined by $(X^\Lambda)'= A^{\Lambda}{}_{
\Sigma} X^\Sigma$.
In this model one can write down the period matrix, as was shown
in \cite{CDFV}
$$
{\cal N}_{\Lambda\Sigma}(X)=
(s-\bar s)
(\Phi_\Lambda\bar\Phi_\Sigma +\bar\Phi_\Lambda\Phi_\Sigma)+\bar s
\eta_{\Lambda\Sigma} \ ,
$$
where $\Phi_\Lambda=\eta_{\Lambda\Sigma}\Phi^\Sigma$, where one can
 raise or lower indices with $\eta$ and
$
\Phi^\Lambda={{ X^\Lambda}\over{\sqrt{ X^\Sigma \eta_{\Sigma\Pi}
\bar X{}^\Pi}}}\ .
$
The real part of the period matrix is
$$
\rm Re \, {\cal N}_{\Lambda\Sigma}(X)=
(s+\bar s)
\eta_{\Lambda\Sigma} \, .
$$
This gives us an information on the gauge coupling for the vector
fields.  In particular, the axion from the $s$ field, the
$C_{(4)}$, couples to $FF*$. It cannot be shifted by a holomorphic
function of $x^2$ which would be required, as shown in
\cite{KalloshHsuProk} to change the K\"ahler potential from $({\rm
Im}x)^2$ to $x\bar x$-type.

In the sector of the gauge fields where the Abelian gauge symmetry
is enhanced to a non-Abelian symmetry, instanton corrections may
lead to terms in the action of the form $e^{-1/g^2_{YM}}$. This in
turn may be understood as coming from a holomorphic superpotential
$W=e^{-a s}$, where $a$ is some constant. Notice that the
definition of the volume-4-form superfield $s$ is such that the
K\"ahler potential has a shift symmetry under the translation of
the D7 brane position, i. e. it depends on $({\rm Im} \,  x)^2$.
The analysis of the special geometry shows that the
non-perturbative potential for the KKLT stabilization can depend
on $e^{-a s}$ but cannot depend on any holomorphic function of $x$
as it follows from the period matrix ${\cal N}$. This accomplishes
the proof of the shift symmetry for the motion of the D7 brane,
both in the K\"ahler potential and in the superpotential.

\section{ Shift symmetries in positions of  D3 branes}

%%%%%%%%%%%%%%%%%%%%%%%%%%%
The  symplectic section and the K\"ahler potential in absence of D7 moduli are given by
\begin{tabbing}
\hskip 9 cm \= \kill \\
\hskip 4 cm $X^- = 1$\,,\> $F_-^X =  -stu +u\,(y^r)^2/2$\,,\\
\hskip 4 cm $X^+ = -tu$ \,,\>$F_+^X = s$\,,\\
\hskip 4 cm $Y^- = -t$\,,\>$F^Y_- = -su$\,,\\
\hskip 4 cm $Y^+ = -u$\,,\>$F_+^Y = -st + (y^r)^2/2$\,,\\
\hskip 4 cm $X^r = y^r$\,,\>$F_r = -u\,y^r$\,.\\
\end{tabbing}
 \begin{equation} K = -\log[-8\,{\rm Im}(u)] - \log [{\rm
Im}(s){\rm Im}(t)+\frac{1}{16}\,({\rm Im}(y)^r\,)^2] \,.
\end{equation}
The shift in D3 position with real parameters $\beta^r$,
\begin{equation}
 (y^r )' = y^r +\beta^r \qquad s'=s \qquad t'=t \qquad u'=u
\label{D3shift}
\end{equation}
can be realized as a lower triangular block form  symplectic matrix:
$${\cal S}^{D3} = \left(\matrix{A & 0\cr C & (A^T)^{-1}}\right)$$
In case of one D3 brane, $r=1$ we find for $A^{\Lambda}{}_{ \Sigma}$
\begin{eqnarray}
&&A\,=\,\,\left(\matrix{1 &0&0&0&0\cr 0
&1&0&0&0\cr 0 &0&1&0&0\cr 0 &0&0&1&0\cr \beta &0 &0&0&1}\right)\,,
\label{AD3}\end{eqnarray}
and for $C_{\Lambda \Sigma}$
\begin{eqnarray}
&&C\,=\,\,\left(\matrix{0 &0&0&-\beta^2/2&0\cr 0 &0&0&0&0\cr 0
&0&0&0&0\cr  \beta^2/2&0&0&0&\beta\cr 0 &0
&0&\beta &0}\right)\,. \label{AD3a}\end{eqnarray}
One can calculate the period matrix and find out the gauge
couplings  for this case. Alternatively, one can study its
symmetry properties using Eq. (\ref{periodtrans}), adapted to our
case.
To simplify things we may look at the imaginary part of the period
matrix, ${\rm Im}\, { \cal N}$.
\begin{equation}
({\rm Im}\, { \cal N})' = (A^T)^{-1} \,{\rm Im}\, { \cal N} A^{-1}
\label{periodtrans2}
\end{equation}
The crucial observation here is that the total period matrix ${
\cal N}_{\Lambda \Sigma}$ can be split into the part which has
terms which interact with the graviphoton vector field and those
which are decoupled. In our basis  this is a split into ${ \cal
N}_{--}, \, { \cal N}_{- i}$ and ${ \cal N}_{ij}$. Here $i$
includes all components but $-$. Using the explicit form of the
matrix $A$ (\ref{AD3}) in Eq. (\ref{periodtrans2}) one can
establish that only the components ${ \cal N}_{--}, \, { \cal
N}_{- i}$ transform, all components ${ \cal N}_{ij}$ are invariant
under the shift. The non-perturbative instanton superpotential
depends only on gauge coupling in the  $ij$ sector of the theory,
graviphoton vector field does not contribute \footnote{At the
point of enhancement of gauge symmetry where the non-Abelian gauge
symmetry arises, the graviphoton does not contribute by symmetry
reasons.  We are grateful to T. Banks  for explaining this.}. Thus
we conclude that the superpotential must be shift invariant when
we use special coordinates for which K\"ahler potential was
invariant from the very beginning.
\section{Conclusion}
The bottom line of this strict special geometry analysis for
cosmological applications is the following. Assuming that the
volume of $T^2$ is the same as the one for $K3$, and that the
dilaton and complex structure are fixed via fluxes as in
\cite{Giddings:2001yu},
 one finds the K\"ahler potential used in this paper. Using the notation of \cite{KalloshHsuProk} one finds
$K= -3\log [(\rho+\bar \rho  -(\phi+\bar \phi)^2]$ for D3 and $K=
-3\log [(\rho+\bar \rho)] +(S+\bar S)^2/2$ for D7. Also one may
follow ADFT \cite{Angelantonj:2003zx} and introduce the gaugings
which represents turning on fluxes in string theory. The gaugings
lead to a non-vanishing potentials, which may break ${\cal N}=2$
down to ${\cal N}=1$ and stabilize the axion-dilaton and the
complex structure. The gauging (introduction of fluxes) does not
affect the period matrix of vector multiplets. Therefore, as shown
above, the non-perturbative superpotential which may be used for
stabilization must be shift-invariant and is given by $W= W_0+ A
e^{-a\rho}$ as in \cite{KKLT}, \cite{KalloshHsuProk}. This is
valid for the case when either D3 is light and D7 is heavy and
only D3 can move or D7 is light and D3 is heavy and only D7 can
move. In both cases the position of the moving brane is a modulus,
since both the K\"ahler potential and the superpotential are shift
invariant.

It is amazing that the proof of the shift symmetry for the
application  to cosmology is based on exactly the same set of
special geometry tools (see eqs.
(\ref{specialcoordinate})-(\ref{duality})), which were used
earlier in studies of  extremal black holes: they explained   the
attractor behavior of the scalars  near the black hole horizon and
the duality symmetries of the black hole entropy formula.

In the setting of this paper  special geometry controls duality
symmetries which include a shift symmetry for the inflaton field.
This may provide a fundamental  basis for the realistic string
cosmology where  almost flat potentials are required for explanation
of the cosmological observations.

We are grateful to C. Angelantonj, T. Banks, S. Ferrara.  S.
Kachru, A. Linde and J. Maldacena for stimulating discussions.
This work  is supported by NSF grant PHY-0244728. J.H. is
also supported by a NSF Graduate Research
Fellowship.

\appendix
\section{Shift symmetries of homogeneous non-symmetric space $L(0, n_7, n_3)$}

Here we consider the general case when both positions of D7 as well as positions of D3 branes are turned on. The section of
the $Sp(2(4+n_7+ n_3),R)$ bundle is
\begin{equation}
X=\left(\begin{array}{c} X^- = 1 \\X^+ = -tu+x^{2}/2\\Y^- =
-t\\Y^+=-u\\X^k = x^k\\X^r=y^r\end{array}\right)
\end{equation}
\begin{equation}
F=\left(\begin{array}{c} F_-^X = s(-tu+x^{2}/2)+u y^2 /2
\\F_+^X=s \\ F_-^Y = -su\\F_+^Y=-st+y^2 /2\\F_k=-sx^k \\F_r=-uy^r \end{array}\right)
\end{equation}
Under $x^k \rightarrow x^k + \alpha^k$ and $y^r \rightarrow y^r + \beta^r$,
\begin{equation}
\Delta X=\left(\begin{array}{c} 0 \\\alpha^k X^k + \alpha^2 X^- /2\\ 0 \\
0 \\ \alpha X^-\\\beta X^-\end{array}\right)
\end{equation}
\begin{equation}
\Delta F=\left(\begin{array}{c}  -\alpha^k F_k + \alpha^2 F_+^X /2 -\beta^r F_r -\beta^2 Y^+ /2\\
0 \\ 0 \\ \beta^r X^r + \beta^2 X^- /2 \\-\alpha F_+^X\\\beta Y^+
\end{array}\right)
\end{equation}
Note that $\Delta F$ has contributions from both $X$ and $F$,
whereas $\Delta X$ involves only $X$. This implies that $B$ in the
transformation matrix above is zero. We can reconstruct $A$, $C$
and $D$ by noting that $X+\Delta X = A X$ and $F+\Delta F = C X +
D F$. Using this, we get that,
\begin{equation}
A =
\left(\begin{array}{cccccc}   1&0&0&0&0&0\\
\alpha^2 /2 & 1 & 0 & 0 & \alpha & 0\\
0&0&1&0&0&0\\
0&0&0&1&0&0\\
\alpha &0&0&0&1&0\\
\beta &0&0&0&0&1
\end{array}\right)
\end{equation}
\begin{equation}
D =
\left(\begin{array}{cccccc}   1&\alpha^2 /2&0&0&-\alpha &-\beta \\
0& 1 & 0 & 0 & 0 & 0\\
0&0&1&0&0&0\\
0&0&0&1&0&0\\
0&-\alpha &0&0&1&0\\
0&0&0&0&0&1
\end{array}\right)
\end{equation}
\begin{equation}
C =
\left(\begin{array}{cccccc}   0&0&0&-\beta^2 /2&0&0\\
0& 0 & 0 & 0 & 0 & 0\\
0&0&0&0&0&0\\
\beta^2 /2&0&0&0&0&\beta\\
0&0&0&0&0&0\\
0&0&0&\beta&0&0
\end{array}\right)
\end{equation}
We need to check that the 10x10 matrix is symplectic which amounts
to the conditions $A^T C = (A^T C)^T$ and $A^T D = I$. This is in fact true so the
transformation is a symplectic one.
As far as the period matrix is concerned, we  need to know only the properties of the gauge fields kinetic matrix of the form $f(z, \bar z)\eta_{AB}$ where the graviphoton has to be excluded. This will provide us the information on the kinetic terms for the vector fields with the enhancement of gauge symmetries to the non-Abelian ones.
\begin{equation}
{\rm Im} {\cal N}_{\Lambda \Sigma}'= {\rm Im} f (D \eta D^T)_{\Lambda \Sigma} \end{equation}
An explicit calculation gives for $D \eta D^T$
\begin{eqnarray}
&&D \eta D^T=\,\,\left(\matrix{\beta^2 &1&0&0&\beta\cr 1
&0&0&0&0\cr 0 &0&0&1&0\cr 0 &0&1&0&0\cr 0 &\beta &0&0&-1}\right)\,.
\end{eqnarray}
which proves that $f'=f$. Therefore the non-perturbative superpotential is shift symmetric.

\end{document}